\documentclass{article}

\PassOptionsToPackage{numbers}{natbib}
\usepackage[final]{neurips_2019}

\usepackage[utf8]{inputenc} 
\usepackage[T1]{fontenc}    
\usepackage{hyperref}       
\usepackage{url}            
\usepackage{booktabs}       
\usepackage{amsfonts}       
\usepackage{nicefrac}       
\usepackage{microtype}      
\usepackage{graphicx}
\usepackage{subcaption}

\usepackage{verbatim}

\title{Estimating localized complexity of \\white-matter wiring with GANs}

\makeatletter
\newcommand{\printfnsymbol}[1]{%
  \textsuperscript{\@fnsymbol{#1}}%
}
\makeatother

\author{%
  Haraldur T. Hallgr\'imsson\thanks{Equal contribution.} \\
  University of California, Santa Barbara \\
  \texttt{hth@cs.ucsb.edu} \\
  \And
  Richika Sharan\printfnsymbol{1}
  \\University of California, Santa Barbara \\
  \texttt{richikasharan@cs.ucsb.com} \\
  \AND
  Scott T. Grafton \\
  University of California, Santa Barbara \\
  \texttt{scott.grafton@psych.ucsb.edu} \\
  \And
  Ambuj K. Singh \\
  University of California, Santa Barbara \\
  \texttt{ambuj@cs.uscb.edu} \\
}

\begin{document}

\maketitle

\section{Introduction}

In-vivo examination of the physical connectivity of axonal projections through the white matter of the human brain is made possible by diffusion weighted magnetic resonance imaging (dMRI)~\citep{basser1994mr,moseley1990diffusion}. Analysis of dMRI commonly considers derived scalar metrics such as fractional anisotrophy as proxies for ``white matter integrity,'' and differences of such measures have been observed as significantly correlating with various neurological diagnosis and clinical measures such as executive function~\citep{grieve2007cognitive}, presence of multiple sclerosis~\citep{cercignani2001mean}, and genetic similarity~\citep{kochunov2015heritability}. 

The analysis of such voxel measures is confounded in areas of more complicated fiber wiring due to crossing, kissing, and dispersing fibers~\citep{volz2018probabilistic}. Recently, Volz et al.~\citep{volz2018probabilistic} introduced a simple probabilistic measure of the count of distinct fiber populations within a voxel, which was shown to reduce variance in group comparisons. We propose a complementary measure that considers the complexity of a voxel in context of its local region, with an aim to quantify the localized wiring complexity of every part of white matter. This allows, for example, identification of particularly ambiguous regions of the brain for tractographic approaches of modeling global wiring connectivity~\citep{maier2017challenge}. 

Our method builds on recent advances in image inpainting~\citep{pathak2016context,Iizuka2017image,yu2018generative}, in which the task is to plausibly fill in a missing region of an image. This task requires both an understanding of the surrounding context of the image, as well as the ability to generate realistic content. Tremendous progress has been made in accomplishing this using deep convolutional neural networks~\citep{krizhevsky2012imagenet} and Generative Adversarial Nets (GANs)~\citep{goodfellow2014generative}. Both of these research areas have known pitfalls which are exacerbated with smaller datasets, as is the case with brain imaging: adversarial examples confound deep learning models~\citep{goodfellow2014explaining}, whereas combating mode collapse in GANs is an active area of research~\citep{srivastava2017veegan}. 

Bayesian deep learning introduces more robust models, especially in the limits of small data. Our proposed method builds on a Bayesian estimate of heteroscedastic aleatoric uncertainty of a region of white matter by inpainting it from its context~\citep{kendall2017uncertainties}. This uncertainty captures the noise inherent in each input data, due for instance to sensor noise or measurement fidelity. We define the localized wiring complexity of white matter as how accurately and confidently a well-trained model can predict the missing patch. In our results, we observe low aleatoric uncertainty along major neuronal pathways which increases at junctions and towards cortex boundaries. This directly quantifies the difficulty of lesion inpainting of dMRI images at all parts of white matter. Previous work has incorporated such uncertainty modeling to improve dMRI super-resolution~\citep{tanno2017bayesian}, though they do not control for higher uncertainty associated with high image intensity.

\begin{figure*}[t]
    \centering
    \begin{subfigure}[t]{0.18\textwidth}
        \centering
        \includegraphics[height=1.3in]{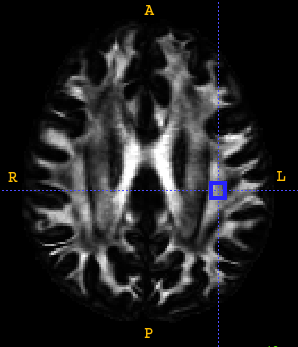}
        \caption{Location of patch.}
    \end{subfigure}
    \begin{subfigure}[t]{0.38\textwidth}
        \centering
        \includegraphics[height=1.3in]{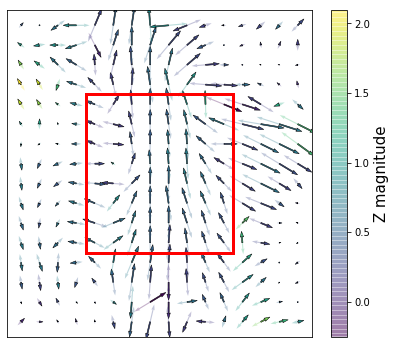}
        \caption{Ground-truth MDA vectors.}
    \end{subfigure}
    \begin{subfigure}[t]{0.38\textwidth}
        \centering
        \includegraphics[height=1.3in]{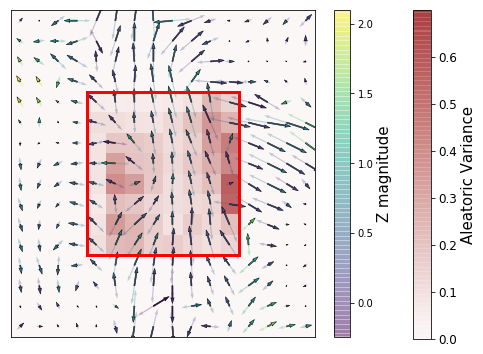}
        \caption{Model output.}
    \end{subfigure}
    \caption{Sample ground-truth (middle) and model output (right) of a patch and its context located within a ``crossing pocket''~\citep{volz2018probabilistic} (left, MDA magnitude weighted background), shown as a 2D projection of a 2D slice of the $16 \times 16 \times 16 \times 3$ tensor. The z-magnitude of the 3D vectors is displayed as a blue to yellow color. The red square delineates the extent of the patch; the exterior context with the interior patch masked is the input to the generator. The estimated aleatoric variance is shown as a red overlay, with more red corresponding to less confident predictions.}
    \label{fig:intro}
\end{figure*}

\section{Methods}

We consider a vector field model of a diffusion-weighted image, in which the vectors correspond to the largest direction of multidirectional anisotrophy (MDA)~\citep{tan2015multi}. We seek to regress a $(n \times n \times n \times 3)$-sized patch, corresponding to all 3D vectors encompassed by a cube of side length $n$, from only its $(2n \times 2n \times 2n \times 3)$-sized context, in which the center patch has been masked. Refer to Fig.~\ref{fig:intro} for an example from the test dataset. We use the notion of voxel coherence~\citep{hallgrimsson2018spatial} to model the intrinsically symmetric nature of the MDA peaks, which a vector field model otherwise would not support.

Selecting an appropriate value of $n$ involves a trade-off between localizing the complexity well for small values of $n$ versus a more challenging but holistic notion of wiring complexity for larger values. In this study we selected $n=8$, corresponding to a patch size of $10\times 10\times 10\,$mm$^3$ due to the 1.25x1.25x1.25 mm$^3$ spatial resolution of the diffusion volumes. 

We train a conditional GAN model to accomplish this~\citep{mirza2014conditional}. Given the context of a patch, with the patch itself masked out with zeros, it generates the most likely estimate of the patch along with its confidence in each output dimension of that prediction. These predictions are not only encouraged to match the known ground-truth patch, but additionally to depict a realistic enough wiring pattern such that the discriminator cannot discern ground-truth from generated patches.

The discriminator optimizes a typical cross entropy loss, but we adopt the following for the generator,

\begin{equation}
\label{eq:loss}
    \mathcal{L}_{G_\theta} = \log D(\hat{p} | C) + \sum_{v \in V} \frac{1}{2} \hat{\sigma}_v^{-2} d(p_v, \hat{p}_v) + \frac{1}{2} \log \hat{\sigma}_v^2
\end{equation} 

where $p$ is the patch under consideration and $C$ its associated context, $\hat{p}$ and $\hat{\sigma}^2$ the model's predicted patch and aleatoric variance, $v$ indexes the set of voxels $V$ within the patch, $d(x, y) = \min(\|x-y\|_2, \|x+y\|_2)$ the symmetric L2 norm~\citep{hallgrimsson2018spatial}, and $D(\hat{p} | C)$ the probability that the discriminator outputs for the generated patch given its context. Eq.~\ref{eq:loss} equally encourages the generator to balance ``fooling'' the discriminator and predicting the ground truth patch, with the aleatoric variance $\hat{\sigma}^2$ being a learned loss attenuation of that prediction which is penalized for large uncertainty.

We adopt a coarse-to-fine architecture following Yu et al.~\citep{yu2018generative}, which predicts an initial crude patch (optimized solely for reconstruction loss) before the final output which optimizes Eq.~\ref{eq:loss}. To stabilize the convergence of the adversarial model, we apply one-side label smoothing~\citep{salimans2016improved}. The model was trained with a batch size of 32 on two GeForce RTX 2080s for 19 hours, or seven thousand epochs.

\section{Data}

We apply our model on the same dataset as Volz et al.~\citep{volz2018probabilistic} of 630 scans collected as part of the Human Connectome Project~\citep{van2013wu}. In particular, the minimally processed HCP images were identically reconstructed using generalized Q-sampling imaging~\citep{yeh2010generalized} and registered to a multimodal template using symmetric group-wise normalization~\citep{avants2010optimal}. The single largest multidimensional anisotrophy (MDA) peak was extracted from each voxel, an example of which can be seen in Fig.~\ref{fig:intro}.

We uniformly at random assigned each scan to one of training ($n=442$), validation ($n=94$), or test ($n=94$). During each epoch of training, we uniformly at random selected a different white matter patch from each scan. Hyperparameters were tuned based on their performance on the validation set, and all results presented are from the test set.

\section{Results}

\begin{figure*}[t]
    \centering
    \includegraphics[width=0.85\textwidth]{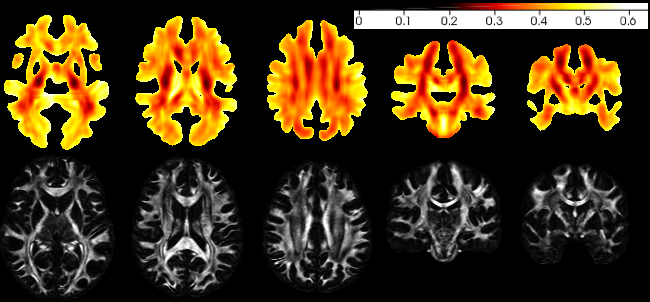}
    \caption{The aleatoric standard deviation normalized by average magnitude (top) and respective max MDA-weighted map (bottom), with a patch size of $k=8$. Brighter variance values correspond to higher uncertainty estimated by the model, as indicated by the colorbar (top-right). A minimum magnitude threshold has been applied to the aleatoric variance to avoid spurious results.}
    \label{fig:results}
\end{figure*}

The model was evaluated on every other voxel to obtain average aleatoric variance and symmetric $L2$ errors, with results linearly interpolated between. These two values correlate highly (Pearson coefficient of $r=0.831$, $p<10^{-10}$), indicating the aleatoric variance is well calibrated. Fig.~\ref{fig:results} shows the average aleatoric coefficient of variation (the square root of the variance normalized by MDA magnitude) across all test scans for each patch for select slices. This measure controls for greater model error, and thus uncertainty, associated with regions of higher MDA image intensity.

The normalized aleatoric uncertainty is lowest within major fiber bundles (e.g., dark regions in center of slices in Fig.~\ref{fig:results}), which are relatively uniform structures, though greater near junctions or crossings of those bundles. It is greatest where white matter fibers disperse into gray matter. We observe a high uncertainty, relative to nearby regions, near the ``crossing pockets'' (seen in Fig.~\ref{fig:intro}, and Fig.~\ref{fig:results} top row, middle column) identified by Volz et al.~\citep{volz2018probabilistic} as having multiple distinguishable orientations.

\section{Discussion}

These results are an average across a population, and as such assumes a certain homogeneity in wiring patterns, or the complexity thereof, across individuals. However, we observe relatively high uncertainty in regions which are known to be the most heterogeneous within this same population~\citep{volz2018probabilistic}. Our future work will aim to better quantify and understand heterogeneity, especially in the presence of complex wiring patterns. A current limitation of our model, and in our future understanding of these heterogeneous regions, is that this study only considered the single largest MDA peak per voxel. 
We observed difficulty in obtaining stable convergence of adversarial models when further peaks were considered, but this is not a fundamental limitation.

Lesioned brains offer particular challenges for group analysis, as the lesions are confounders for group normalization~\citep{crinion2007spatial}. The method presented here could naturally be extended to ameliorate such challenges. We intend to further study the effects of the patch size $n$ to assess its usefulness for larger lesions, as well as its effects on complexity analysis such as that considered in this study.

\section{Acknowledgements}

We are grateful to the maintainers of various software packages that made this study possible, including ITK-SNAP~\citep{py06nimg}, DSI Studio (http://dsi-studio.labsolver.org), and ANTs~\citep{avants2009advanced}. This work was supported by NSF award 1817046, Contract W911NF-09-0001, and Cooperative Agreement W911NF-19-2-0026 with the Army Research Office of the Army Research Laboratory.

\bibliography{refs.bib}

\newpage
\appendix
\section{Architecture}
\begin{figure}
    \centering
    \includegraphics[width=1\textwidth]{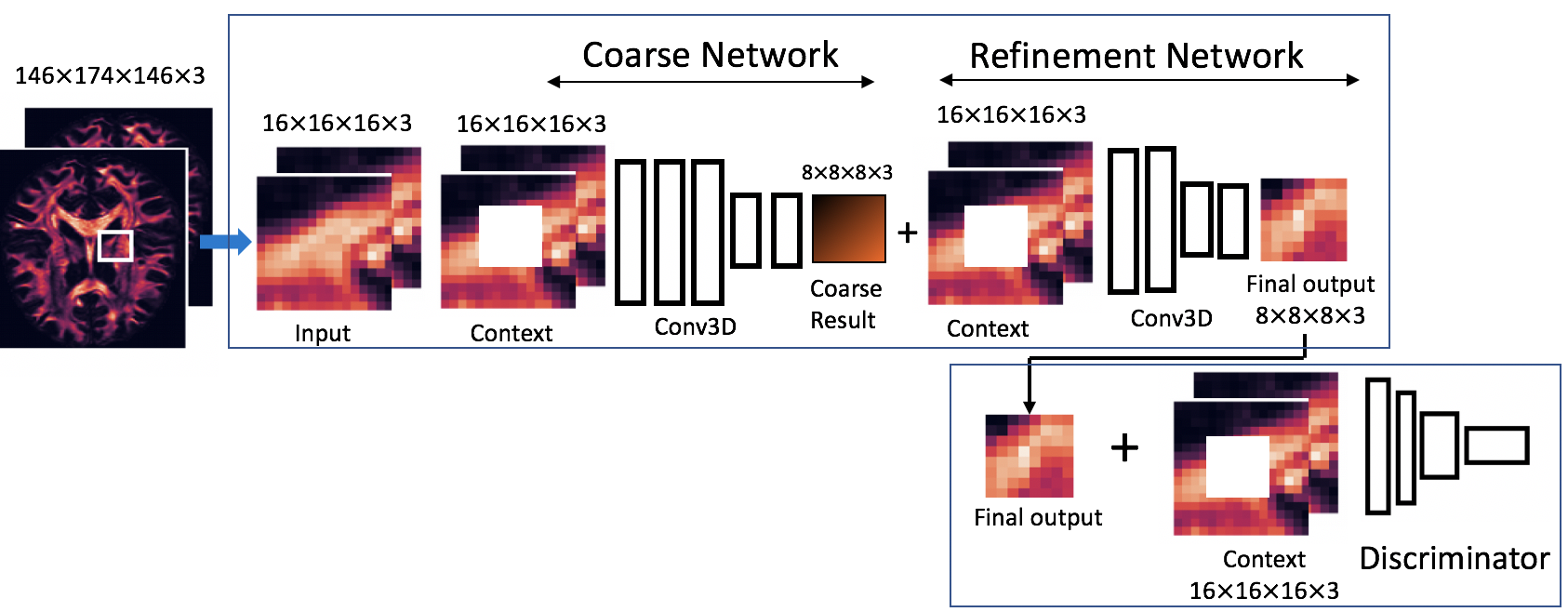}
    \caption{GAN Architecture}
    \label{fig:architecture}
\end{figure}

We describe the GAN generator and discriminator architecture in detail in this appendix, as seen in Fig.~\ref{fig:architecture}. For simplicity, we use the following abbreviations for the convolutional layers: K (kernel size), C (channel number), S (stride size), D (dilation rate), and BN (Batch normalization). All convolutional layers have a Leaky ReLU activation with $\alpha = 0.3$ and same padding. The outputs of both generators are clipped to five standard deviations of the mean, as derived from the training set. The kernel weights of the convolutional layers were initialized with a He normal distribution~\cite{he2015delving}, while all other weights were sampled from a Glorot uniform.

\textbf{Discriminator:} K3,C32,S1,BN - K3,C64,S2,BN - K3,C128,S2,BN - K3,C128,S1,BN - K3,C128,S1,BN - dense layer with one channel and sigmoid activation.

\textbf{Generator - Coarse network:} K3,C128,S1 - K3,C128,S1 - K3,C128,S1 - K1,C64,S2 - K1,C3,S1.

\textbf{Generator - Fine network:} K3,C128,S1 - K3,C128,S1 - K3,C128,S1 - K3,C128,S1,D2 - K3,C128,S1,D4 - K1,C64,S2 - K1,C3,S1.

The last layer of the refinement network (K1,C3,S1) is duplicated, once each for the predicted output and the aleatoric variance, receiving the same input but otherwise updated independently using Eq.~\ref{eq:loss}. 

\end{document}